\documentclass[12pt]{article}
\usepackage{pic03}
\usepackage{hyperref}
\usepackage{url}
\usepackage{graphicx}

\begin{document}

\title{\bf Direct Dark Matter Searches}
\author{
Wolfgang Seidel        \\
{\em Max-Planck-Institut f\"ur Physik, F\"ohringer Ring 6, 80805 M\"unchen, Germany}}
\maketitle

%
% photograph of author
%  This is where we will insert a photograph. To see what it would look like,
%  uncomment the following lines.
%
%\begin{figure}[h]
%\begin{center}
%
% include photograph for proceeding version
%
%\includegraphics
%[height=4.5cm]{einstein.eps}
%
% insert a fixed vertical spacing instead for the ArXiv preprint
%
\vspace{4.5cm}
%
%\end{center}
%\end{figure}

\baselineskip=14.5pt
\begin{abstract}
The elucidation of the nature of dark matter is one of the challenging tasks in astroparticle physics.  A brief overview on the different methods to search directly for dark matter in form of Weakly Interacting Massive Particles (WIMPs) is given.
\end{abstract}
\newpage

\baselineskip=17pt

\section{Introduction}
Since 1933 when Fritz Zwicky found the first evidence for the existence of dark matter, astronomers and cosmologists became more and more convinced that a large fraction of the mass of our universe is existing in an unknown form of matter - the so called dark matter. In spite of the prominent role this dark matter seems to play on all scales from galaxies up to the structure of the entire universe,  it was up to now not possible to detect it directly nor to elucidate its nature. Dark matter searches are therefore among the hottest topics in astroparticle physics. Among the many theories proposing possible dark matter candidates, so called Weakly Interacting Massive Particles (WIMPs) are presently most favoured. These weakly interacting particles with masses of several GeV  form halos around galaxies. Due to their huge abundance, their gravitational mass dominates the galaxy and may therefore explain the rotational velocity distribution of spiral galaxies. The existence of WIMPs would also imply physics beyond the standard model of particle physics. In a sub-class of supersymmetric theories the neutralino would be a good candidate. 

The detection reaction in direct WIMP searches is the elastic scattering of a WIMP off a nucleus giving the nucleus a recoil energy of up to several ten keV.

\section{General Considerations of Direct Detection Experiments}
The density of our galaxy`s WIMP halo at the position of the earth is estimated to be about $0.3\,GeV/cm^3$. Their velocity distribution is roughly a Boltzmann distribution with a mean velocity of about $270\,km/s$.  Interacting in a detector, WIMPs create nuclear recoils. The energy spectrum of the recoils can be roughly described by an exponential rise towards  low energies \cite{smith}\cite{jung}  . The shape of the recoil spectrum depends, among other parameters, on the WIMP mass, the mass of the target nucleus and the velocity distribution. For typical WIMP parameters, an interaction rate in the order of $1\,count/ day$ or even less per kilogram of detector mass is expected. Obviously, identifying this low rate constitutes an extremely demanding challenge for direct detection experiments.Therefore, background reactions coming from radioactivity in the detector and its surrounding as well as from cosmic rays have to be reduced  or shielded  to at least the same level as the WIMP interaction rate. Meeting these stringent demands requires a continuous development and learning process.

In order to be able to see most of the recoil events the energy threshold of the detector should  be as low as possible. A good energy resolution is  needed  to measure the shape of the recoil spectrum and to determine the origin of the background by gamma and alpha spectroscopy.

Due to the expected shape of the WIMP recoil spectrum the bins right above the detector threshold have the strongest influence on the detector sensitivity. A good monitoring of the gain stability and trigger efficiency is therefore essential for giving reliable exclusion limits or claiming long term modulation effects. 

 There are several ways to positively identify  a WIMP signal. First one can check the shape of the recoil spectrum and its dependence on the mass of the target nucleus.  The  influence of the target nucleus on the interaction rate may be used to elucidate the WIMP`s nature. In addition the rate and the spectrum should exhibit an annual modulation \cite{freese}\cite{spergel}. This modulation is caused by a modulation of the WIMP velocity distribution on earth due to the movement of the earth around the sun.  However, this modulation effect is  only in the order of a few percent of the total WIMP interaction rate. To identify a signal positively as a WIMP signal, several of these should be observed in a consistent manner.  

A nuclear recoil in a solid state detector can be detected in several ways:
Most of the energy of a recoil will create phonons and will therefore result in a small temperature rise of the detector. This temperature rise is detected in cryogenic detectors.
Only about 10\% of the recoil energy is used for creating electron-hole-pairs. These can be measured as a charge signal, for example in high purity Germanium detectors.
If the detector material is a scintillator, about 1\% of the recoil energy may be converted into scintillation light.

On the other hand , however, ionisation and scintillation yields are much higher if the primary interaction takes place with an electron, creating an electron recoil instead of a nuclear recoil. This is the case for all background events created by photons an electrons.  Usually they  constitute the major background component. By measuring two excitation channels simultaneously ( e.g. phonon plus ionisation, phonon plus scintillation or ionisation plus scintillation) this background component can be strongly suppressed. New detectors just coming online use this background suppression technique and will boost the experimental sensivity in the near future by several orders of magnitude. They are limited by the background created by the elastic scattering of neutrons at nuclei, as this type of interaction also creates nuclear recoils indistinguishable from those created by WIMPs. For these experiments neutron shielding is an important issue.

\section{Experiments}

In the following section different experimental techniques are shortly  described and their advantages and disadvantages outlined.
\subsection{Ionisation Detectors}
First dark matter searches were made Germanium detectors. These searches were a byproduct of the search for  neutrinoless double beta decay in germanium. Germanium diodes exhibit an excellent energy resolution and long term stability. Since they are the standard detector for gamma spectroscopy, there exists a lot of experience in running and in industrial production of these detectors. From the radioactivity point of view they can be made very clean and have presently the lowest primary background of all solid state detectors. The restriction  of this technology to Germanium as detector material and the lack of a background suppression may be seen as a drawback for future dark matter searches. Especially the activation of the detectors by cosmic rays during the production  may become a limiting factor which can be overcome only by strong financial and technological efforts. Nevertheless there is a strong potential left in the next future. Dark matter searches with Ge-diodes are, for example, the Heidelberg Moscow Experiment\cite{hdm}, Igex\cite{igex}, GTF\cite{gtf} and in the future Majorana\cite{majorana}.

\subsection{Scintillation Detectors}

\subsubsection{\it Liquid Xenon}

Liquid Xenon is a good scintillator which has the capability to discriminate pulses caused by nuclear recoils (neutrons or WIMPs) from that of electron recoils (gamma and beta background) by exploiting their different time constants of the scintillation pulses. The distributions of these two event classes are, however, strongly overlapping. Therefore a background suppression can be made only on  a statistical rather than an event-by-event basis. This statistical method is used by the Zeplin I dark matter search. The limiting factors  are systematic errors which may be reduced by dedicated calibration runs. Probably due to the difficulties and ambiguities of these runs the limits given by ZEPLIN I \cite{zeplini} are still labeled preliminary. Future liquid Xenon experiments are ZEPLIN II and ZEPLIN III \cite{zeplinxxx}, potentially leading to ZEPLIN-MAX.

\subsubsection{\it NaJ}

NaJ scintillation detectors are standard detectors for gamma spectroscopy and therefore a well established and commercially available technique. The difficulty in using these detectors for dark matter search is the necessity to obtain an extremely low threshold. The DAMA collaboration has been running about 100kg of NaJ counters with a threshold of 2\,keV for more than seven years collecting 107731\,kg\,days of data \cite{dama}. They clearly find an annual modulation in the event rate in their lowest energy bins. Because they cannot attribute this modulation to any other origin,  they interpret it to be caused by an annually modulated WIMP signal hidden in their background. This interpretation is in partial contradiction but at present not fully excluded by other experiments as EDELWEISS or CDMS.

\subsection{Cryogenic Calorimeters}
Cryogenic calorimeters consist of dielectric or superconducting crystals cooled to temperatures below 100\,mK. A suitable thermometer is attached to these crystals. Even with crystal masses of a few hundred grams an energy deposit  of a few keV results in a measurable temperature rise (some $10^{-6}\,^\circ K$) of the thermometer. Detectors of this type were  developed during the last 15 years and there are still a lot of ongoing  improvements. The main advantage of this detectortype is the extremely low threshold down to about 600\,eV for nuclear recoilsand the possibility to use a large variety of detector materials. On the other hand , however, the complexity of producing and running these devices is certainly a disadvantage. An excellent background discrimination can be obtained by the simultaneous measurement of heat and ionisation or heat and scintillation light as described in the next sections.
Pure cryogenic dark matter searches are CRESST I\cite{cressti}, the Tokyo LiF / NaF setup\cite{tokyo}, Rosebud\cite{rosebudi}, Cuoricino and in future Cuore\cite{cuore}.

\subsection{Phonon plus Ionisation}
This measurement technique, developed by the CDMS and EDELWEISS collaborations, is in simple words the combination of a Germanium detector with a cryogenic calorimeter. Since background events (produced by photons and electrons) create much more charge than nuclear recoil events of the same energy, the ratio of charge signal to phonon signal (the phonon signal is independent of the type of event) can be used to classify the event type and therefore to suppress beta and gamma background on an event by event basis.
The fact that the sensitivity of the CDMS\cite{cdms} and EDELWEISS \cite{edelweiss} dark matter searches with a few ten kilogram days of data is competitive to detectors with masses up to hundred kilograms running for several years shows the advantage of this detector concept. CDMS is just starting a physics run in their deep underground lab in the Soudan mine which has the capability for making a big step forward.
 
\subsection{Phonon plus Scintillation}
The simultaneous detection of scintillation light and phonons in cryogenic calorimeters using scintillating absorber crystals can give a similar background suppression as the simultaneous measurement of ionisation and light. This method was developed by the CRESST collaboration for CRESST II\cite{cresstii}. Very recently it was shown\cite{coron}  that a large variety of scintillating crystals can be used. This gives this method a big advantage in identification of WIMP signals. The experiments using this technique are CRESST II and Rosebud.

\section{Conclusion}
Dark matter searches are one of the hot topics of astroparticle physics but are experimentally very challenging.  A key issue of every experiment is to obtain a background rate as low as possible. In the future this can only be obtained by experiments having the capability of a background suppression on an event by event basis. In the next few years  improvements  of about two orders of magnitude in sensitivity seem feasible, which will clarify some still ongoing discussion and cover already a substantial part of the theoretically predicted parameter rage for  neutralino  dark matter. If we are lucky, next generation experiments  might perhaps even find the dark matter.

\end{document}